# Observation of a Two-Dimensional Electron Gas at the Surface of Annealed SrTiO$_3$ Single Crystals by Scanning Tunneling Spectroscopy

R. Di Capua[1,2], M. Radovic[1,3], G. M. De Luca[1], I. Maggio-Aprile[4], F. Miletto Granozio[1], N. C. Plumb[5], Z. Ristic[1,4], U. Scotti di Uccio[1,6], R. Vaglio[1,6] and M. Salluzzo[1]

[1]CNR-SPIN, Complesso Universitario Monte S. Angelo, Via Cintia I-80126 Napoli, Italy

[2]Dipartimento S.pe.S., Università degli Studi del Molise, Via De Sanctis, I-86100 Campobasso, Italy

[3]LSNS - EPFL, PH A2 354 (Bâtiment PH) Station 3 CH-1015 Lausanne, Switzerland

[4]Département de Physique de la Matière Condensée, University of Geneva, 24 Quai Ernest-Ansermet, CH-1211 Geneva 4, Switzerland

[5]Swiss Light Source, Paul Scherrer Institut, CH-5232 Villigen PSI, Switzerland

[6]Dipartimento di Scienze Fisiche, Università "Federico II" di Napoli, Via Cintia I-80126 Napoli, Italy

ABSTRACT

Scanning tunneling spectroscopy suggests the formation of a two dimensional electron gas (2DEG) on the TiO$_2$ terminated surface of undoped SrTiO$_3$ single crystals annealed at temperature lower than 400 °C in ultra high vacuum conditions. Low energy electron diffraction indicates that the 2D metallic SrTiO$_3$ surface is not structurally reconstructed, suggesting that non-ordered oxygen vacancies created in the annealing process introduce carriers leading to an electronic reconstruction. The experimental results are interpreted in a frame of competition between oxygen diffusion from the bulk to the surface and oxygen loss from the surface itself.

PACS: 68.47.Gh, 73.20.-r, 87.64.Dz

# 1. Introduction

The formation of a two dimensional electron gas (2DEG) at the interface between $SrTiO_3$ (STO) and $LaAlO_3$ (LAO) has attracted huge interest since its discovery [1] because of the intriguing functionalities related to this phenomenon, including high electron mobility [2], large magnetoresistance [3], and superconductivity [4]. The mechanism at the origin of the 2DEG formation is still the subject of much debate. Electronic reconstruction and charge transfer at the interface, in the "polar catastrophe" scenario, is arguably the most likely cause of the 2DEG formation [1,5], but oxygen vacancies and structural/chemical rearrangements at the interface have also been considered [6-8]. The overall picture is clouded by the presence of several complicating factors, such as the possibility to have n- or p-type interfaces (according to the termination [5]), or the mixing between STO and LAO layers at the interface itself [9].

While the formation of the 2DEG at the STO/LAO interface is now a well accepted fact, it is not so obvious whether bare STO can exhibit the same phenomenon at its interface with vacuum. Without the presence of the polar LaO and $AlO_2$ planes, an electronic reconstruction is not expected at the STO (100) surface, since the atomic planes are formally neutral. However, due to the partially covalent character of the cation-oxygen bonding, contrary to what the nominal charges of ions suggest, the SrO and $TiO_2$ surfaces of STO are weakly polar, and thus they are subject to structural and possible electronic instabilities [10]. In addition, from an experimental point of view, while in LAO/STO a 2DEG at the interface is protected by the LAO overlayer, a free STO surface is in general unstable, and experiments in this sense must overcome this difficulty.

Very recent papers [11,12] demonstrated the formation of a 2DEG on the surface of in-situ cleaved STO by angle resolved photoemission spectroscopy (ARPES). The authors cleaved STO crystals in-situ at cryogenic temperatures in vacuum, so that the obtained "fresh" surface was preserved. Oxygen vacancies, formed in the crystal fracturing [11] or triggered by exposure to ultraviolet light [12], have been indicated as responsible for the 2DEG formation and confinement at the surface. Very interestingly, some of the electronic properties of the fresh cleaved surface are similar to that one of the LAO/STO interface, and in particular they exhibit a similar splitting between 3dxy and 3dxz/3dyz bands [13,14]. On the other hand, scanning tunneling spectroscopy (STS) measurements on STO crystals heated at high temperature (of the order of 800 – 1200 °C) revealed gap-like features [15] (annealing is necessary to give conducting properties to a stoichiometric, insulating, STO crystal, making it suitable for STS measurements). It has not yet been clarified whether the apparently opposite results between photoemission and tunnel analyses should be ascribed to possible differences in the techniques and in the physical phenomena that they probe (so that they cannot really be compared), or instead are related to truly intrinsic differences in the samples due to the specific preparation procedures.

Here, we studied, by STS measurements, the density of states (DOS) around the Fermi level of STO single crystals, for samples treated with different annealing procedures to produce the oxygen-deficient state. We found that thermal treatments on well-ordered $TiO_2$ surfaces at relatively low temperatures (250 °C) in ultra high vacuum conditions create a surface layer hosting a 2D electron system. On the other hand, a treatment at higher temperature recovers the insulating surface state.

## 2. Experimental

STO crystals with (001) surface cut were purchased from SurfaceNet GmbH. Their surfaces were treated by a standard etching (in HF solution) [16] and annealing procedure (950 °C, 1 hour in flow of oxygen) to realize a pure $TiO_2$-terminated surface. The samples were then placed in a ultra high vacuum (UHV) chamber (base pressure ≈ $5x10^{-11}$ mbar) and treated in oxygen atmosphere (0.1 mbar, 350°C for one hour) before being in-situ transferred to the UHV scanning tunneling microscope (STM) chamber (further details on samples preparation methods in our labs can be found in Ref. [17]). The structural quality of the surface was checked by reflection high energy electron diffraction (RHEED), showing the very clear 1x1 diffraction pattern of the non-reconstructed (100) STO surface. This treatment made the STO crystals insulating and transparent, as expected for stoichiometric STO without oxygen vacancies: any attempt to establish a tunnel current between the tip and the surface of such samples failed. Atomic force microscope (AFM) images revealed, on these insulating samples, the presence of atomically flat terraces (the mean roughness on the terrace being less than 0.1 nm, Fig. 1a), separated by single unit cell steps. The thermal treatments described below resulted instead in conducting surfaces, that we explored by scanning tunneling microscopy and spectroscopy (STM/STS). The measurements were realized at room temperature by using a commercial Omicron VT-AFM scanning tunnelling microscope equipped with W or PtIr tips. Depending on the particular measurement, the tip to sample bias voltage was set in the range of 1-2 V, with a tunnel current of 0.1 – 0.7 nA (tunnel resistance in the range 1-20 GΩ).

## 3. Results and discussion

We started with annealing at relatively low temperatures of 250-350 °C and pressures of about $10^{-11}$ mbar. After this treatment, a visual inspection reveals that samples remain transparent. However, as mentioned above, their surfaces were conducting as revealed by the absence of charging effects in low energy electron diffraction (LEED) and by STM measurements. Fig. 1b-c show an STM image of a sample annealed at 250 °C for 12 hours in UHV conditions (Sample A), together with a height profile. These data demonstrate that high surface quality was preserved on the annealed crystals. The clear 1x1 LEED pattern obtained from such surface demonstrates that no reconstruction takes place (Fig. 2a). A thermal treatment at a slightly higher temperature of 350 °C (Sample B) does not change the morphology, and the LEED pattern remains 1x1. On the contrary a 2x1 reconstruction is observed on $TiO_2$ terminated STO annealed at 900 °C (Sample C, Fig. 2b). These samples are bulk conducting, due to the oxygen loss from the whole crystal.

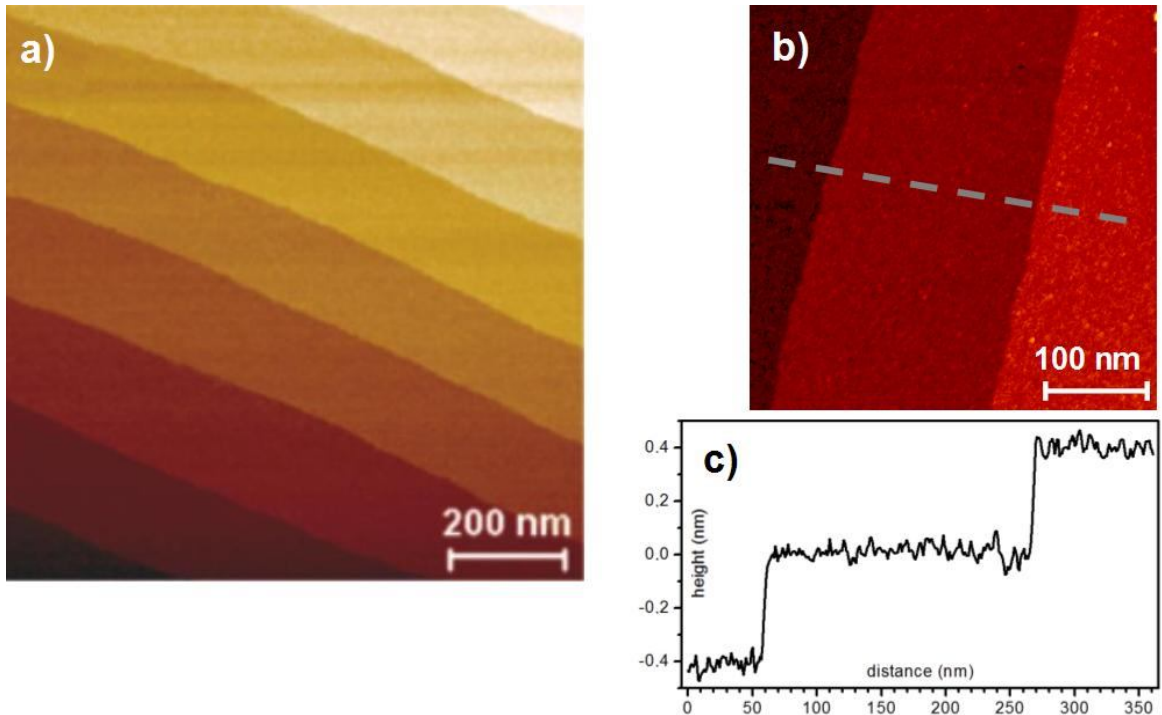

Fig. 1: a) Non-contact AFM topography (1 μm x 1 μm) on a stoichiometric STO crystal, showing the terrace structure of the surface. b) STM topography (400 nm x 400 nm) on Sample A (annealed at 250 °C), and c) height profile along the shown dashed line.

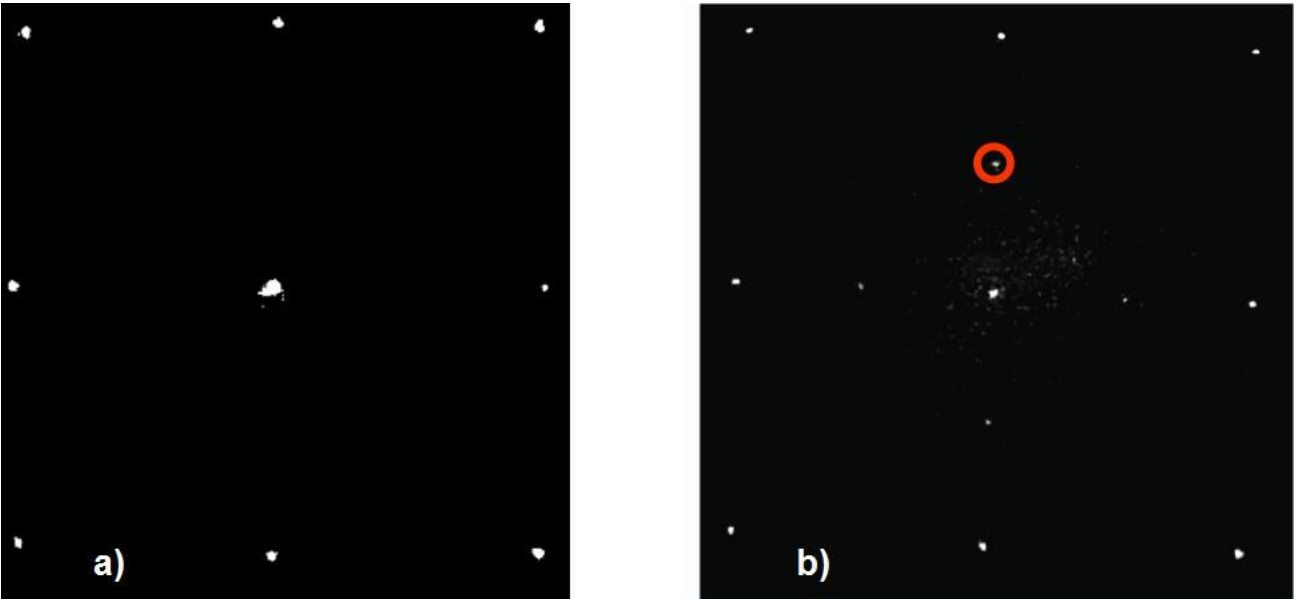

Fig. 2: a) LEED image on Sample A exhibits a 1x1 pattern; the same result is obtained on Sample B, proving that the surface is not reconstructed at the lowest annealing temperatures. b) On Sample C, the LEED pattern shows a 2x1 reconstruction of the surface.

Tunnel current and differential tunnel conductance (by mean of a standard lock-in technique) vs. bias voltage (dI/dV curves) were recorded on Samples A, B, C (Fig. 3a and Fig. 3b). The curves were acquired over a grid of points on several scanning regions, in order to improve statistics and to check the homogeneity of the sample surface.

The 2x1 reconstructed Sample C exhibits a strong asymmetry between occupied (negative bias) and unoccupied (positive bias) states, and a quite flat near-zero signal below the Fermi energy. This behavior reflects the strongly asymmetric DOS of bulk STO and its surfaces also in the presence of oxygen vacancies. The doping of electrons into the empty 3d titanium band of STO moves the Fermi level towards the edge of the conduction band. The valence band is instead 3.2 eV below the conduction band, so that no states are available in the gap region, and the tunnel conductivity is very low. The dI/dV characteristics on Sample C are in agreement with the previously observed behavior of reconstructed STO [15,18], indicating a non-fully conducting surface. On the other hand, the low-T annealed Sample A (250 °C) shows very different STS spectra, which are unusual in many respects. The dI/dV spectra show a non-zero conductivity at the Fermi level, an almost symmetric behavior at low bias for occupied and unoccupied states, and a V-shaped background. The spectra also show other distinctive characteristics like the features around ±250 meV, reproducibly observed on the whole surface, and a depletion of states (a dip) around the Fermi level. Annealing at slightly higher temperature of 350 °C, as in Sample B, produces a gap-like feature.

This occurrence was observed by STS on non-reconstructed $TiO_2$ terminated STO [15,18], suggesting a more insulating character of this surface, in agreement with *ab-initio* calculations [15]. In order to highlight the main features of the electronic properties of the conducting Sample A, we estimated the local density of states (LDOS) through the standard normalization procedure established in [19], i.e. LDOS=(dI/dV)/(I/V). This procedure was already applied to STO surfaces by Tanaka et al. [18]. The LDOS, shown in Fig. 3c, is characterized by pronounced peaks around ±250 meV and a finite zero bias conductivity together with a strong depletion around the Fermi level. We can also notice a shoulder around -100 meV (Fig. 3c), more or less pronounced depending on the location, but observed in all spectra, as shown by the sequence of (dI/dV)/(I/V) spectra taken along a line in a map (Fig. 3d). At higher bias-voltage, the spectra have an almost flat background that for occupied states exhibits a very broad peak starting at -1.0 eV with its broad maximum at about -1.7eV. The sequence of spectra also confirms a substantial reproducibility of the spectroscopic features as a function of the position, thus demonstrating that they are distinctive features of the low temperature annealed Sample A.

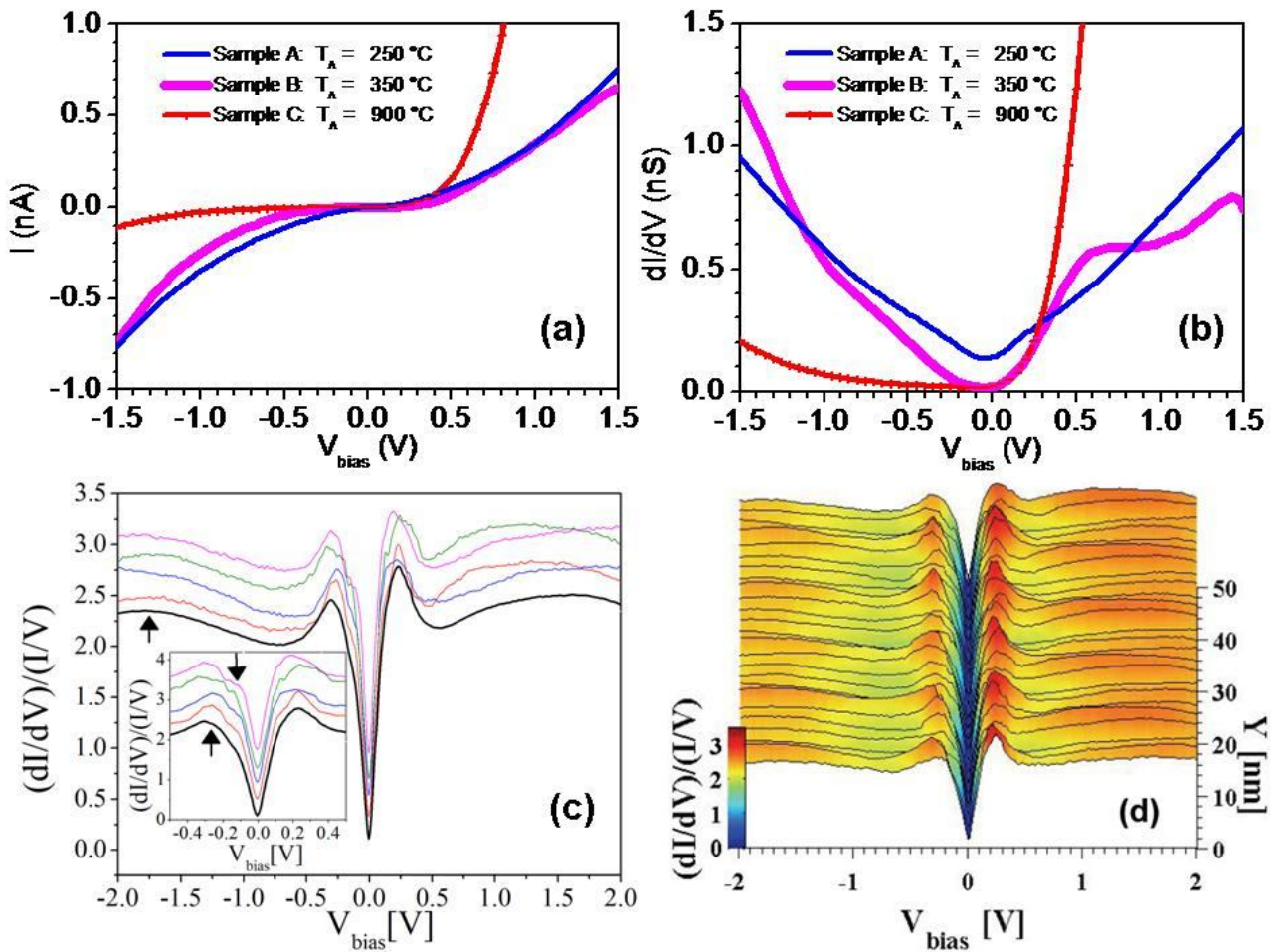

Fig. 3: a) Tunneling I-V and b) dI/dV recorded on Samples A, B, and C ($T_A$ = annealing temperature). Tunneling curves were recorded at several locations on the surfaces forming a grid of point for each measured crystal, and the displayed curves are an average of all such single measurements. c) Tunneling DOS estimation using the normalization (dI/dV)/(I/V) for Sample A; the inset shows the presence of the structures at -100 mV and -1-1.3 eV in each location. d) The single DOS curves for Sample A on a scanning line, demonstrating the reproducibility of the measurements and the substantial homogeneity of the surface.

These results should be compared with theoretical calculations and recent ARPES measurements performed on freshly cleaved STO surfaces, which demonstrated the existence of a 2D electron system [11,12]. One of the signatures of the 2DEG state is the presence of split sub-bands below the Fermi level, which are not expected for STO bulk samples. The sub-bands were created by the confinement of electrons, inside a thin surface layer, in a potential well arising from the bending of the conductance band close to the surface. Santander et al. [11] were able to resolve the split sub-bands, originated by the overlapping of neighboring titanium 3d orbitals. The lowest one extends its DOS below the Fermi energy [11,12], with its maximum between 200 meV and 300 meV (depending on the crystalline momentum). Higher in energy, other 3d sub-bands were also identified, between 100 meV and the Fermi level. Our findings on Sample A, annealed at 250 °C, are consistent with the results obtained by ARPES, suggesting the formation of a 2DEG. In fact, we observe a non-zero DOS below the Fermi level. Furthermore, the pronounced peak at -250mV andthe shoulder at about -100 mV are, in our opinion, related to the split sub-bands observed by ARPES. Differently from the ARPES case, our data allow us to probe unoccupied states showing the presence of similar structures above the Fermi level.

An interesting observation is the apparent re-opening of a gap-like structure, indicating an evolution back to the insulating state, when the annealing temperature is increased. As demonstrated in Ref. [12], the features of the 2DEG on the STO surface are independent of the STO bulk properties over a large range of bulk carrier density $n_{3D}$, from less than $10^{13}$ cm$^{-3}$ (insulating STO) to $10^{20}$ cm$^{-3}$ (strongly doped). The experimental setup does not grant us quantitative control of $n_{3D}$. However, we can qualitatively conclude that annealing is not able to produce such a high bulk doping level for samples annealed below 400 °C, since the samples remain transparent [11]. Thus, we expect that the observed evolution vs. annealing temperature is essentially due to the behavior of the surface 2DEG and of the density of oxygen vacancies at the surface. This is especially true for samples annealed at the lowest temperatures.

Such non-monotonic behavior might be due (at least in the low temperature annealing regime) to the competition of the thermal activated oxygen fluxes from the bulk to the surface and from the surface to the vacuum. If the latter is characterized by a lower activation energy, at a relatively low temperature it predominates on the other one, realizing a depletion of the oxygen from the surface.

Increasing the temperature, the flux towards the vacuum saturates while the flux from the bulk to the surface becomes comparable, and the global result is the oxygen refilling of the surface. Some preliminary data from x-ray photoemission spectroscopy (XPS) confirm a lower surface vacancies amount in samples annealed at very high temperature (through a lower photoemission signal coming from spectroscopic features attributed to $Ti^{3+}$). The simple proposed mechanism can simply explain the presence of an "optimal" temperature for the stabilization of the 2DEG shown by STS spectra, as well as the oxygen vacancies trends observed by XPS. In this frame, the 2DEG state could disappear just because of the decreasing of free carriers on the surface layer, but also the slighter band bending effect [11] due to the smaller difference between the vacancies concentration between the bulk and the surface can contribute. Of course, we cannot exclude a role of structural reconstruction (actually observed in Sample C), vacancies ordering, or Sr migration, in determining the disappearance of the 2DEG; these effects play a major role likely at the highest temperature.

## 4. Conclusions

In conclusion, we measured the tunnel DOS of $TiO_2$-terminated STO single crystals after annealing in UHV at different temperatures. In the resulting non-insulating state, the presence of a non-zero DOS close to the Fermi level was observed for both empty and occupied states. This finding is in agreement with very recent ARPES measurements on in vacuum cleaved STO, which showed the presence of a 2DEG at the surface. Furthermore, it confirms that the oxygen vacancies, which are the consequence of sample annealing, are responsible for the formation of such 2DEG in the case of bare STO. The presence of an optimal temperature for the observation of the effect could be explained in terms of competition between the oxygen diffusion from the bulk to the surface and the oxygen loss from the surface to the vacuum, which constitutes a further indication of the strong role played by the oxygen vacancies. Further increasing of the annealing temperature determines the disappearance of the 2DEG, and it is possible that more effects besides the oxygen refilling of the surface are responsible of this behavior.

## Acknowledgments

The research leading to these results has received funding from the European Union Seventh Framework Programme (FP7/2007-2013) under grant agreement N. 264098 - MAMA.